\documentclass[
    ,final            
,numberedheadings 
  ]
  {aipproc}

\layoutstyle{8x11double}

\usepackage{graphicx}
\usepackage{subfig}
\usepackage{epstopdf}
\usepackage{dcolumn}
\usepackage{amsmath}

\begin{document}

\title{Towards Determining the energy of the UHECRs observed by the ANITA detector}


\author{Konstantin Belov for the ANITA collaboration}{
  address={UCLA Department of Physics and Astronomy / 154705, 
475 Portola Plaza, 
Los Angeles, CA 90095, USA}
}

\begin{abstract}
The Antarctic Impulsive Transient Antenna (ANITA) is a balloon-borne radio experiment designed to discover ultra-high energy cosmic neutrinos. The ANITA detector has completed one prototype and two full-scale flights above the Antarctic continent. Two direct and fourteen reflected cosmic ray events of ultra-high energy were observed during the first full scale flight and several others in the second flight. We present a Monte Carlo technique and analysis developed to determine the energy of the primary cosmic ray particles from the ANITA data.
\end{abstract}

\maketitle


\section{Introduction}

Cosmic ray particles are constantly bombarding the earth's atmosphere. The energy of some of the cosmic rays can exceed $10^{20}$ eV \cite{PhysRevLett.10.146}, making them the carriers of valuable information about the Universe and particle interactions at ultra-high energies. These ultra-high energy cosmic rays (UHECRs) have been the object of intensive study since the 1938 discovery by Pierre Auger of extensive air showers, cascades of secondary particles in the atmosphere caused by UHECRs. Despite significant progress in research, the origin, acceleration mechanism and chemical composition of UHECRs are still unknown. The flux of the cosmic rays falls rapidly with energy, making direct observations of UHECRs impossible.

Indirect observations employ the earth's atmosphere as a giant calorimeter to observe secondary particles on the ground using muon counters, or to observe air fluorescence or Cherenkov light produced by the extensive air showers using special UV light sensitive cameras. The latter observations can only be conducted during moonless nights, reducing the duty cycle of the detector to less than 10\%. A hybrid technique combines air fluorescence detectors and ground counters, improving data statistics as well as arrival direction resolution, primary particle energy and the depth of the shower maximum ($X_{max}$) reconstruction. This latter parameter of air shower development in the atmosphere can be
used to obtain chemical composition and particle cross-sections at
ultra-high energies \cite{PhysRevLett.104.161101, PhysRevLett.104.091101, Belov2006197}.  The ground array and air fluorescence techniques have advanced significantly over the last several decades, but radio detection still has several significant advantages, including lower deployment and operational costs combined with a 100\% duty cycle. Unfortunately, the interest in radio detection diminished after the initial experiments in the 1960's \cite{allan1970}.  The lack of understanding of the radio production mechanism at that time resulted in somewhat limited success. 

A new breed of ground radio detectors has appeared in the last several years \cite{doi:10.1117/12.551466, Ardouin2005148, 2011arXiv1109.5805C, Fuchs201293} complemented by the Antarctic Impulsive Transient Antenna (ANITA) balloon--borne detector, which reported the first observation of UHECRs in the 200--1200 MHz frequency band in a far zone \cite{PhysRevLett.105.151101}. The ANITA instrument is a self-triggering detector, which puts it apart from ground radio arrays usually located near an existing cosmic ray telescopes that supply trigger information. With the development of Monte Carlo (MC) simulations incorporating precise physics \cite{Ludwig2011438, AlvarezMuniz2012325}, it becomes possible to obtain the energy of the primary cosmic particle as well as information about the air shower development from radio data. This makes radio observations more attractive as a complement to the existing cosmic ray telescopes or as a stand alone independent observation technique. 

A special approach, however, is needed to determine the energy of the primary particle and the air shower development parameters from the radio data. We present a technique to obtain the energy of the UHECRs using MC simulations in conjunction with the total power and spectral density of the observed radio pulses emitted by the extensive air showers.

The radio emission from an extensive air shower increases with primary energy. Due to relativistic amplification, the maximal power will be emitted along the Cherenkov angle of the shower. The proposed technique employs the spectral power density to identify the shower core position with respect to the receiver. To obtain the primary particle energy, one has to combine it with the total broadband power measured by the receiver. Because this technique relies only on measuring the radio emission, it proves to be a powerful method for measuring UHECR energies not only with balloon--borne or space--borne receivers, but also with ground--based detectors. With the ability to reconstruct the Cherenkov ring, the ground--based detectors can also measure the depth of the shower maximum,  $X_{max}$, opening a possibility to study the cosmic ray chemical composition and particle cross-sections at ultra-high energies using the radio data.

\section{ANITA data}

\begin{figure}[b]
\includegraphics[width=0.72\columnwidth]{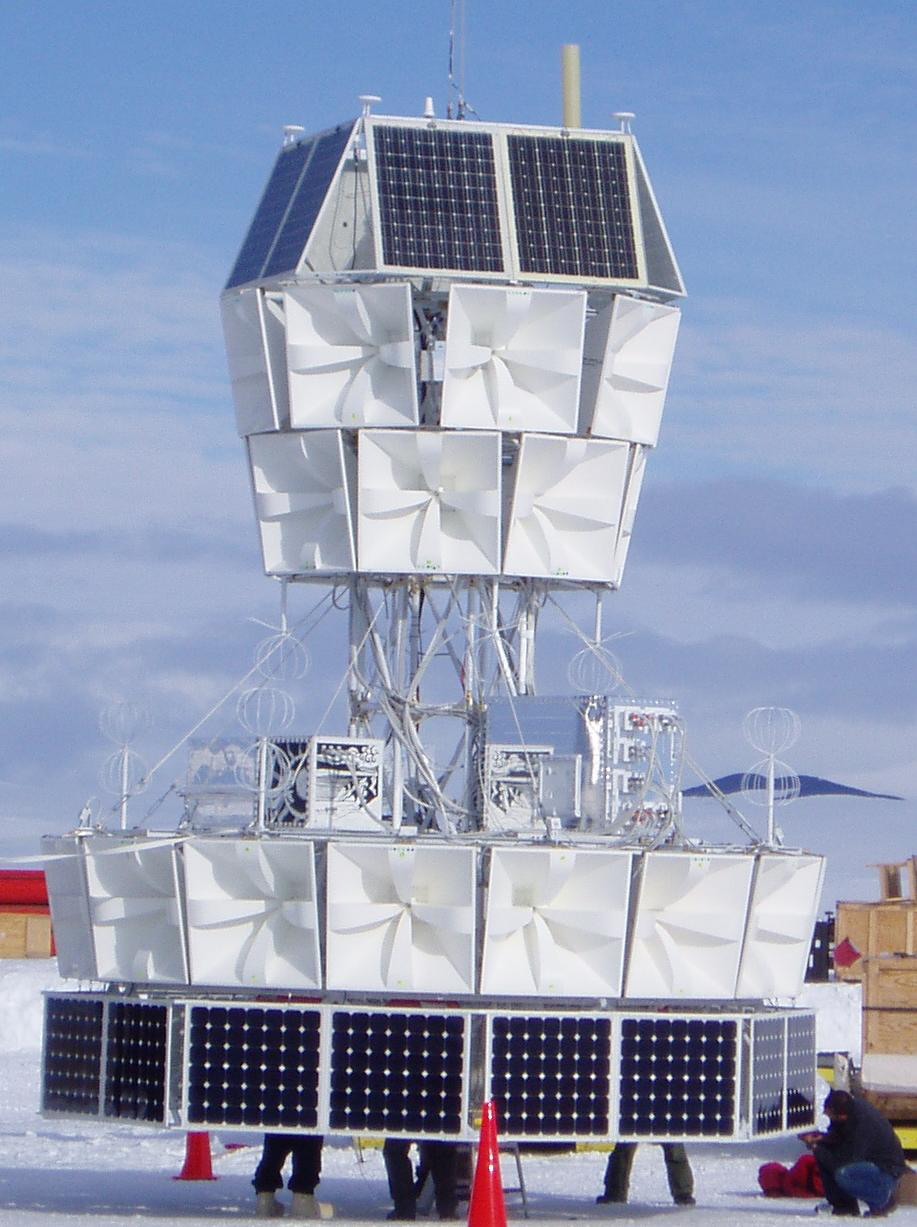}
\caption{ANITA I before the flight. Photo by Jeff Kowalski.}
\label{ANITAIinAntarctica:fig} 
\end{figure}

ANITA is a radio detector designed to observe ultra-high energy cosmic neutrinos \cite{Gorham200910}, see Figure \ref{ANITAIinAntarctica:fig}. An ultra-high energy neutrino interaction in ice will emit a
radio pulse due to the charge asymmetry in the cascade of secondary
particles, the so called Askaryan effect \cite{askaryan:62, askaryan:65}. In case of a neutrino skimming the ice, the radio pulse can be detected from above. Flying some 37 km above the Antarctic ice, the ANITA detector is sensitive enough to pick up weak radio pulses from hundreds of kilometers away and to reconstruct the event location to exclude events with anthropogenic origin. The high altitude provides a large detector aperture while the deep Antarctic ice, with about 2.5 km average thickness, provides a large detector volume. 

\begin{figure}[htb]
\includegraphics[width=\columnwidth]{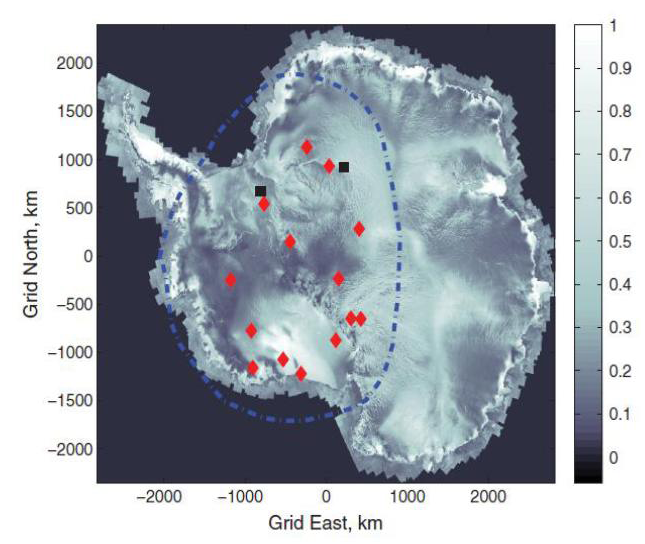}
\caption{Map of locations of detected reflected (red diamond) and direct (black square) UHECR events superimposed on a microwave backscatter amplitude map of Antarctica (Radarsat) within ANITA field of view (dash-dotted line)  \cite{PhysRevLett.105.151101}. For the direct events the closest approach to the surface of the RF signals is shown. The scale on the right is relative amplitude of the backscatter signal.}
\label{anita16events:fig} 
\end{figure}

The ANITA prototype flew in the 2003--04 Antarctic season, followed by the ANITA I flight in 2006--07 and the ANITA II flight in 2008--09. The third ANITA flight is scheduled for the 2013--14 Antarctic season.  The two ANITA flights did not find ultra-high energy neutrinos above the expected background, and a limit on the neutrino flux is published \cite{PhysRevD.82.022004}. The neutrino signal is expected to have a vertical polarization due to the event geometry. However, 16 isolated horizontally polarized events were identified in ANITA I data, see Figure \ref{anita16events:fig}.

\begin{figure}[b]
\includegraphics[width=\columnwidth]{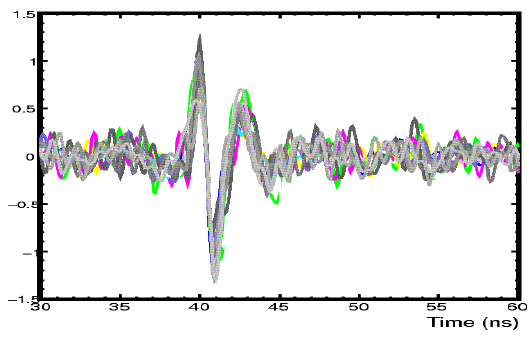}
\caption{Waveforms of 14 ANITA horizontally polarized reflected events, shifted and plotted on the same scale. The instrument response is deconvoluted. Plot by S. Hoover.}
\label{anita16pulses:fig} 
\end{figure}

\begin{figure}[t]
\includegraphics[width=0.95\columnwidth]{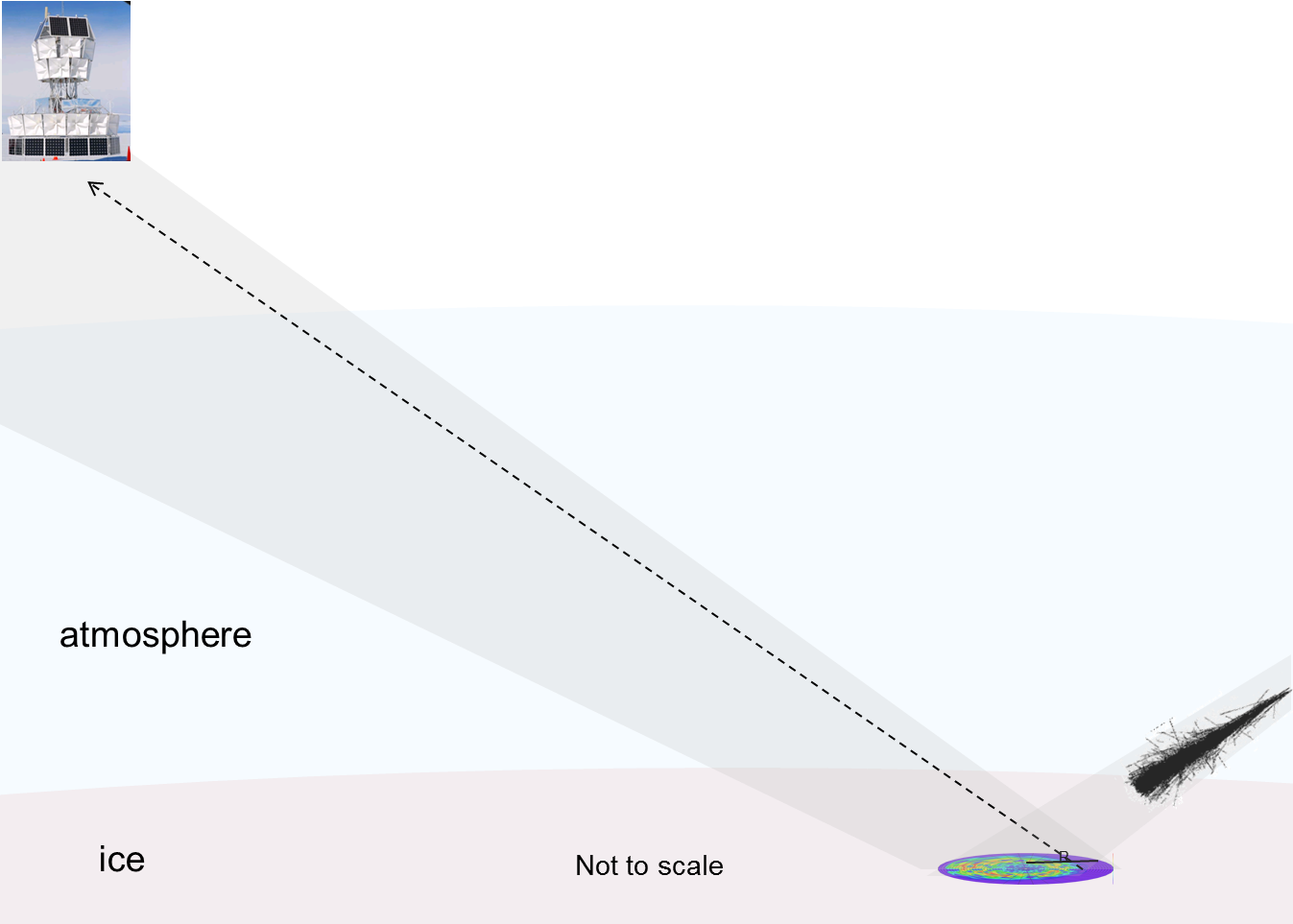}
\caption{ANITA UHECR detector concept.}
\label{ANITACRConcept:fig} 
\end{figure}

These events cannot be associated with ultra-high energy neutrino interactions, and are caused by UHECRs. The 14 events reconstructed on the ice have the same electric field polarity, see Figure \ref{anita16pulses:fig}, while the two events reconstructed above the horizon have opposite polarity,
which is consistent with the geosynchrotron mechanism of radio emission. The radio signal from the 14 extensive air showers, beamed up front into a cone, is reflected from the ice surface and the polarity of the electric field is flipped. An additional argument supporting the geosynchrotron mechanism as cause for the radio signal is that while the ANITA detector did not trigger on the weak vertically polarized content of these events, the residual electric field in the vertical polarization of each event correlates with that expected at the reconstructed event location \cite{PhysRevLett.105.151101}. In fact, ANITA is a sensitive UHECR detector with a very large aperture, see Figure \ref{ANITACRConcept:fig}.
To obtain the energy of the UHECRs from the ANITA data we have to use MC
simulations, taking into account radio frequency (RF) signal propagation in the air,
reflection from the rough ice surface and instrument response.

\section{Monte Carlo simulations}

\begin{figure}[htb]
\includegraphics[width=\columnwidth]{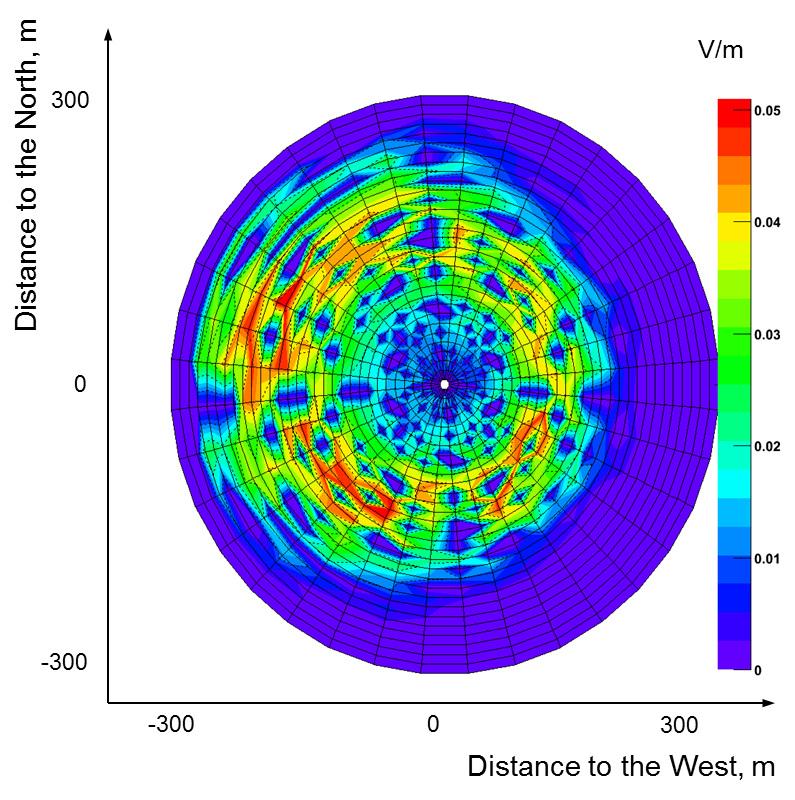}
\caption{Horizontal polarization component of the electric field on the ice surface at the time of maximum electric field magnitude for a simulated ANITA event. Coordinates are relative to the shower core.}
\label{rf_footprint:fig} 
\end{figure}

CORSIKA code version 6.960 \cite{CORSIKA:98} was used for the extensive air shower simulations for this work, and the RF emission from the air shower was calculated using the CoREAS plug-in for CORSIKA \cite{Huegearena2012}. 
The simulations were performed for proton primary particles at different energies from $10^{17}$ to $10^{21}$ eV using the actual event geometries reconstructed by the ANITA detector. The atmospheric parameters were chosen for the South Pole January atmosphere, and the magnetic field at the event locations were used. The MC simulations also take into account the effects of the Fresnel reflection, roughness, antenna response and the RF propagation to the payload as well as the realistic index of refraction, changing with the atmospheric depth. The electric field intensity on the surface of the ice for one of the simulated ANITA events is shown on Figure \ref{rf_footprint:fig}.
As expected for the real data in case of the index of refraction greater than one, the simulation shows that the radio emission forms a ring, actually an ellipse, on the ice surface, corresponding to the Cherenkov cone projection on the ground. For observers inside the ring, the time is reversed, that is, the observer sees the end of the shower first. For observers located somewhere on the ring, the shower develops instantaneously and the emitted signal is amplified due to relativistic effects. The shower core is in the focus of the ring, and the ring geometry depends on the air shower zenith angle and $X_{max}$. 

\section{Energy of the UHECRs}

\begin{figure}[t]
\includegraphics[width=\columnwidth]{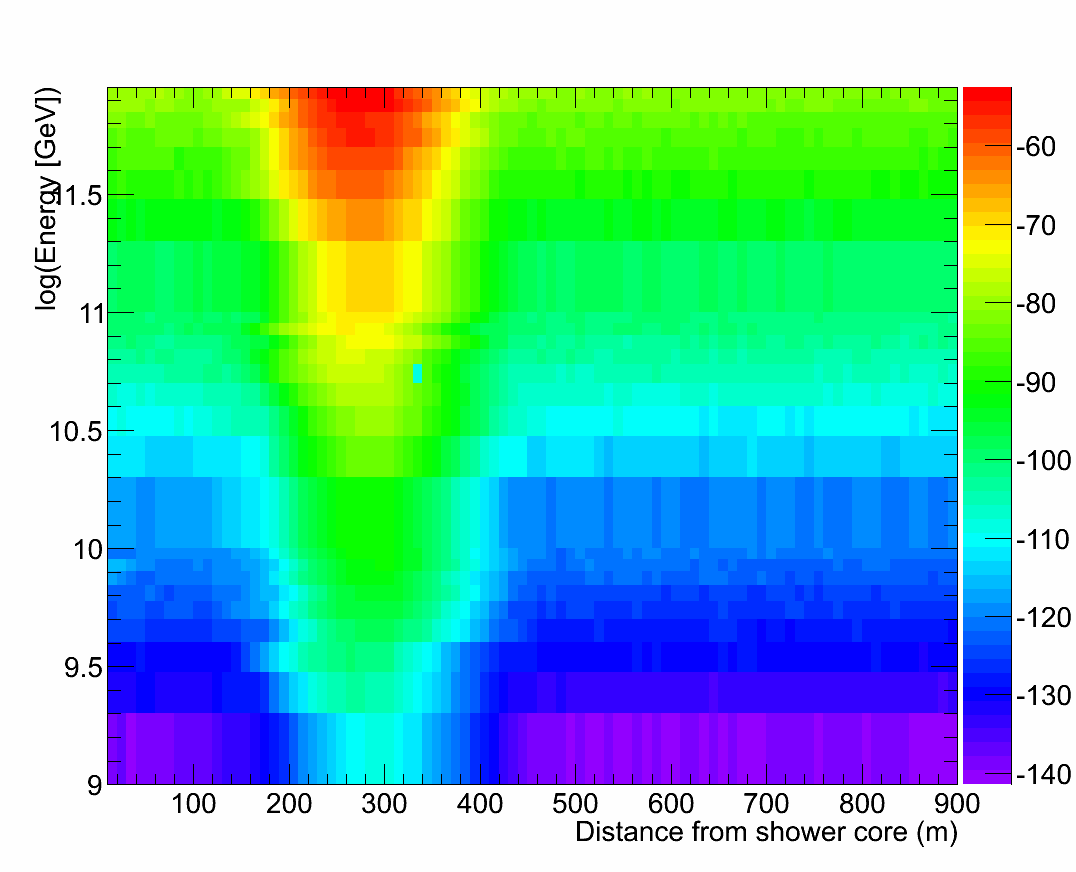}
\caption{Total power in 200--1200 MHz band in units of dBm received by the detector for a series of simulated events of different primary energy. An actual ANITA event geometry was chosen for this simulation. Signal propagation in the atmosphere, ice reflection effects and instrument response are accounted for.}
\label{totalpower:fig} 
\end{figure}

The ANITA detector sees the signal reflected from a location on the Cherenkov ring. Because the detector does not see the signal from the whole ring, the total power of the received radio pulse alone cannot be used to determine the energy of the primary cosmic ray particle. This is true not only for a balloon--borne detector like ANITA, but also for a ground array of radio antennas. Indeed, an observer on the Cherenkov ring can see a very strong signal while an observer inside or outside of the ring will only register noise for the same cosmic ray event. Figure \ref{totalpower:fig} shows the signal intensity received by the payload as a function of primary particle energy and radial distance from the shower core for a series of simulated events of different primary energy. The geometry of one of the 14 ANITA cosmic ray events was chosen for this simulation.

Using only the total power of the received radio signal to determine the energy results in a large uncertainty, because the received signal may have been reflected from any location on the ring.
While there is an obvious ``azimuthal'' coordinate degeneracy of the reflection point seen by the ANITA detector, it is possible to measure the radial distance from the shower core. The signal fluctuations along the Cherenkov ring are significantly less than those in the radial direction and can be included into the systematic error. 

\begin{figure}[tbh]
\includegraphics[width=0.9\columnwidth]{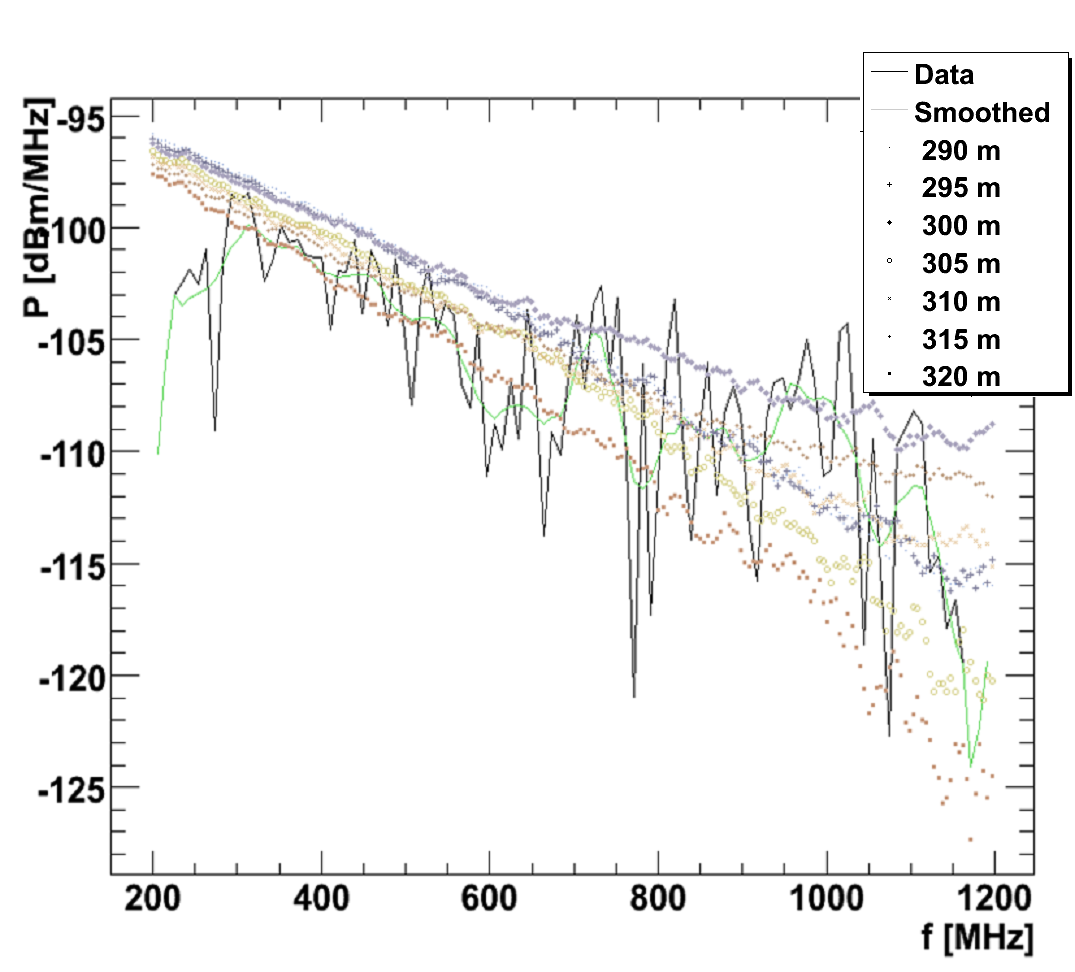}
\caption{Power spectrum density of the radio emission from the simulated air shower and the ANITA event (black line). The smoothed ANITA data (green line) is also shown.}
\label{psd:fig} 
\end{figure}

In addition to finding the total power, a broadband measurement like that
made by ANITA can also determine the power spectrum of the received radio
impulse. The spectrum can be used to determine the reflection point radial distance from the shower core, the $x$-axis coordinate on the plot in Figure \ref{totalpower:fig}, thus limiting the uncertainty on the primary particle energy. Figure \ref{psd:fig} shows the power spectrum density for the same simulated event at different distances from the shower core, together with the actual event data recorded by the ANITA detector. Note the 5 m scale used for the simulation.

\begin{figure}[b]
\includegraphics[width=\columnwidth]{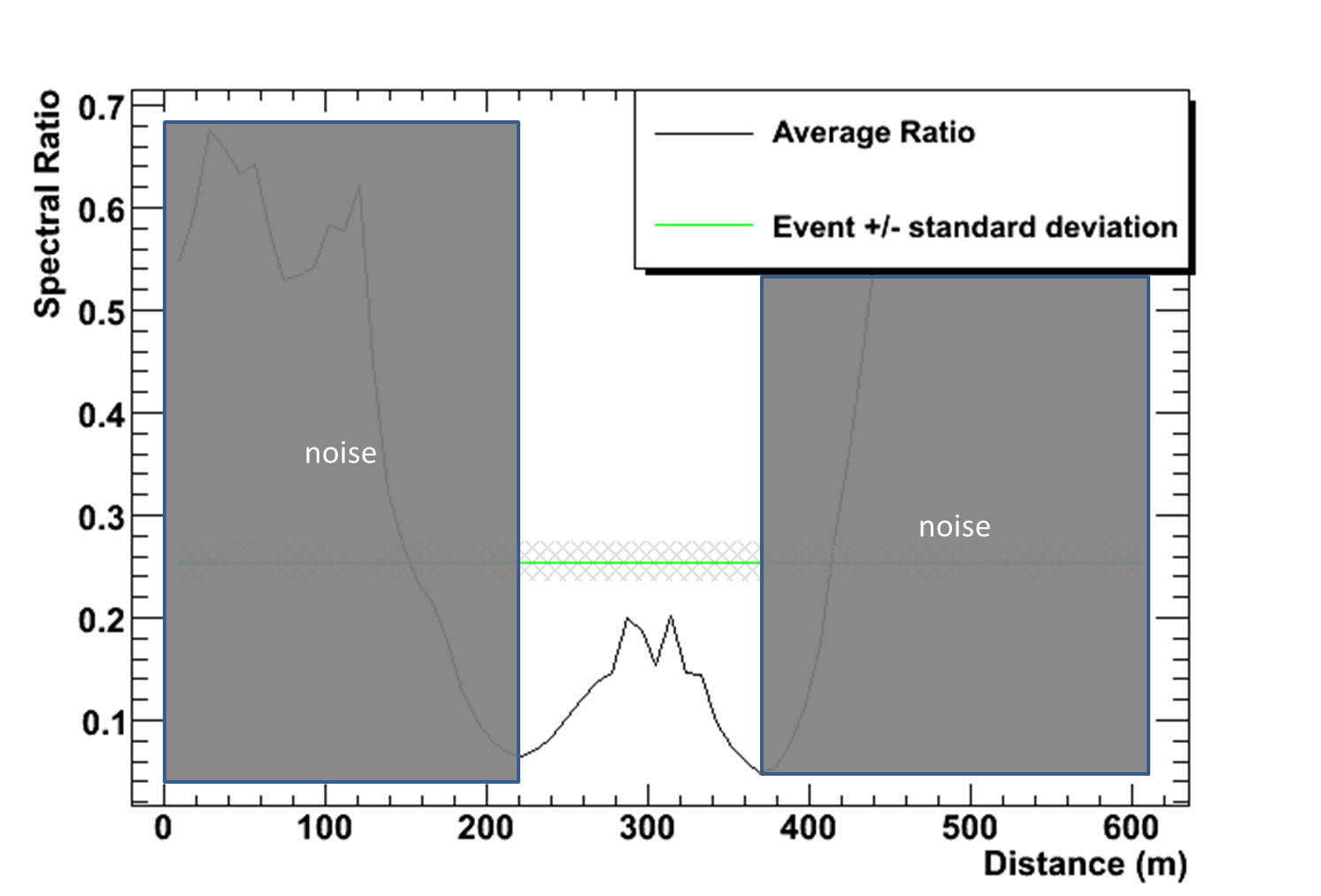}
\caption{Spectral ratio $R$ for the ANITA event (green line) and the simulated event (black line) at different distances from the shower core. There is no signal far from the Cherenkov ring (0--220 m and more than 370 m from the shower core). Those regions are excluded from the analysis.  Because of the noise floor at about -120 dBm/MHz, 20\% of the total power is subtracted from the signal at the higher frequency band for the real data. The MC data does not include noise.}
\label{SpectralRatio:fig} 
\end{figure}

The power spectrum for the simulated data and ANITA events is quantified using a spectral ratio, $R=P_2/P_1$, where $P_1$ is the total power in the 350--600 MHz and $P_2$ is the total power in the 650--850 MHz frequency range, respectively. It is also possible to use the spectral index for this purpose, but for noisy radio signals the spectral ratio is usually more stable. The MC simulations have shown that, as expected,  the spectral ratio $R$ does not depend on the primary particle energy, but changes significantly with the distance from the shower core, reaching its maximum in the middle of the Cherenkov ring, see Figure \ref{SpectralRatio:fig}.
The maximum of the spectral ratio corresponds to the hardest radio spectrum observed from the air shower. On the Cherenkov ring the signal appears coherent at all frequencies, but the coherence diminishes faster for higher frequencies as the observer moves away from the middle of the Cherenkov ring, leading to a softening of the spectrum. The radial distance from the shower core is fixed based on the spectral ratio analysis, which significantly reduces the primary particle energy uncertainty, see Figure \ref{totalpower:fig}. The total power of the ANITA event, of which the geometry was used for this simulation, is --85.69 dBm. Combining the total power of the radio signal, received by the payload, with the limits on the distance from the shower core, obtained from the spectral analysis of the radio signal, one can determine the energy of the primary cosmic ray particle. The uncertainty on the energy measurement depends on several factors including the uncertainties on the signal reflection from the ice, spectrum measurements and the systematic error on the radio signal measurement.

It should be noted that it is not surprising that the ANITA detector sees the signal reflected at a distance from the focus that is centrally located within the Cherenkov ring. The detector was optimized for ultra-high energy {\emph {neutrino}} observations, with the trigger sensitive to very short radio pulses that have broadband and relatively flat spectra. This trigger bias makes the detector sensitive to the middle region of the Cherenkov ring only, and pushes the energy of the observed UHECRs lower by reducing the detector trigger aperture.

\section{Discussion}

The radio spectra from extensive air showers produced by UHECRs provide the additional information needed to measure the energies of cosmic rays. For a balloon-borne or a space-borne radio detector the spectrum reduces the energy measurement uncertainty, putting radio observations on a par with the air fluorescence and the ground counter techniques. 

The technique presented here can also be employed by ground arrays. For a ground radio array the spectral measurements allow for the full reconstruction of the Cherenkov ring using a smaller number of surface radio antennas than those needed in absence of the spectral measurements.  Indeed, the width, size and overall geometry of the Cherenkov ring on the ground varies widely, depending on the frequency range as well as on the extensive air shower zenith angle and the depth of the shower maximum. The dependence on $X_{max}$ opens an opportunity for studies of the cosmic-ray chemical composition and strong interaction particle cross-sections. However, a large array of densely spaced surface antennas is needed to reconstruct the Cherenkov ring, while still covering an acceptable range of air shower parameters with meaningful statistics.  The spectral measurements provide the information about the location of the observer relative to the shower core, effectively reducing the antenna density requirements, and, thus, the total number of antennas needed and the cost of the array. This opens a possibility of building radio arrays that complement the existing detection techniques, or even operate as stand alone UHECR observatories.


\begin{theacknowledgments}
The author wants to express a special thanks to Tim Huege and Marianne Ludwig (Karlsruhe Institute of Technology) for providing the CoREAS simulation code and to David Urdaneta and David Saltzberg (UCLA) for  contributing to this work.  
\end{theacknowledgments}



\bibliographystyle{aipproc}   

\bibliography{KBelovTowardsANITAEnergies}

\def\url#1{}
\begin{thebibliography}{18}
\expandafter\ifx\csname natexlab\endcsname\relax\def\natexlab#1{#1}\fi
\providecommand{\enquote}[1]{``#1''}
\expandafter\ifx\csname url\endcsname\relax
  \def\url#1{\texttt{#1}}\fi
\expandafter\ifx\csname urlprefix\endcsname\relax\def\urlprefix{URL }\fi
\providecommand{\eprint}[2][]{\url{#2}}

\bibitem[Linsley(1963)]{PhysRevLett.10.146}
J.~Linsley, \emph{Phys. Rev. Lett.} \textbf{10}, 146--148 (1963).

\bibitem[Abbasi et~al.(2010)]{PhysRevLett.104.161101}
R.~U. Abbasi, T.~Abu-Zayyad, M.~Al-Seady, M.~Allen, J.~F. Amman, R.~J.
  Anderson, G.~Archbold, K.~Belov, et~al., \emph{Phys. Rev. Lett.}
  \textbf{104}, 161101 (2010).

\bibitem[Abraham et~al.(2010)]{PhysRevLett.104.091101}
J.~Abraham, P.~Abreu, M.~Aglietta, E.~J. Ahn, D.~Allard, I.~Allekotte,
  J.~Allen, J.~Alvarez-Mu{\~n}iz, et~al., \emph{Phys. Rev. Lett.} \textbf{104},
  091101 (2010).

\bibitem[Belov(2006)]{Belov2006197}
K.~Belov, \emph{Nuclear Physics B - Proceedings Supplements} \textbf{151}, 197
  -- 204 (2006), ISSN 0920-5632.

\bibitem[Allan et~al.(2007)]{allan1970}
H.~R. Allan, R.~W. Clay, and J.~K. Jones, \emph{Nature} \textbf{227}, 1116 --
  1118 (2007).

\bibitem[Horneffer et~al.(2004)]{doi:10.1117/12.551466}
A.~Horneffer, T.~Antoni, W.~D. Apel, F.~Badea, K.~Bekk, A.~Bercuci,
  M.~Bertaina, H.~Bluemer, et~al., \enquote{LOPES: detecting radio emission
  from cosmic ray air showers,} in \emph{SPIE Proceedings}, 2004, vol. 5500,
  pp. 129--138.

\bibitem[Ardouin et~al.(2005)]{Ardouin2005148}
D.~Ardouin, A.~Belletoile, D.~Charrier, R.~Dallier, L.~Denis, P.~Eschstruth,
  T.~Gousset, F.~Haddad, et~al., \emph{Nuclear Instruments and Methods in
  Physics Research Section A: Accelerators, Spectrometers, Detectors and
  Associated Equipment} \textbf{555}, 148 -- 163 (2005), ISSN 0168-9002.

\bibitem[{Corstanje} et~al.(2011)]{2011arXiv1109.5805C}
A.~{Corstanje}, M.~{van den Akker}, L.~{B{\"a}hren}, H.~{Falcke},
  W.~{Frieswijk}, J.~R. {H{\"o}randel}, A.~{Horneffer}, C.~W. {James}, et~al.
  (2011), arXiv:1109.5805.

\bibitem[Fuchs(2012)]{Fuchs201293}
B.~Fuchs, \emph{Nuclear Instruments and Methods in Physics Research Section A:
  Accelerators, Spectrometers, Detectors and Associated Equipment}
  \textbf{692}, 93 -- 97 (2012), ISSN 0168-9002.

\bibitem[Hoover et~al.(2010)]{PhysRevLett.105.151101}
S.~Hoover, J.~Nam, P.~W. Gorham, E.~Grashorn, P.~Allison, S.~W. Barwick, J.~J.
  Beatty, K.~Belov, et~al., \emph{Phys. Rev. Lett.} \textbf{105}, 151101
  (2010).

\bibitem[Ludwig and Huege(2011)]{Ludwig2011438}
M.~Ludwig, and T.~Huege, \emph{Astroparticle Physics} \textbf{34}, 438 -- 446
  (2011), ISSN 0927-6505.

\bibitem[Alvarez-Mu{\~n}iz et~al.(2012)]{AlvarezMuniz2012325}
J.~Alvarez-Mu{\~n}iz, W.~R. {Carvalho Jr.}, and E.~Zas, \emph{Astroparticle
  Physics} \textbf{35}, 325 -- 341 (2012), ISSN 0927-6505.

\bibitem[Gorham et~al.(2009)]{Gorham200910}
P.~Gorham, P.~Allison, S.~Barwick, J.~Beatty, D.~Besson, W.~Binns, C.~Chen,
  P.~Chen, et~al., \emph{Astroparticle Physics} \textbf{32}, 10 -- 41 (2009),
  ISSN 0927-6505.

\bibitem[Askaryan(1962)]{askaryan:62}
G.~A. Askaryan, \emph{JETP} \textbf{14}, 441 (1962).

\bibitem[Askaryan(1965)]{askaryan:65}
G.~A. Askaryan, \emph{JETP} \textbf{21}, 658 (1965).

\bibitem[Gorham et~al.(2010)]{PhysRevD.82.022004}
P.~W. Gorham, P.~Allison, B.~M. Baughman, J.~J. Beatty, K.~Belov, D.~Z. Besson,
  S.~Bevan, W.~R. Binns, et~al., \emph{Phys. Rev. D} \textbf{82}, 022004
  (2010).

\bibitem[Heck et~al.(1998)]{CORSIKA:98}
D.~Heck, J.~Knapp, J.~Capdevielle, G.~Schatz, and T.~Thouw,
  \emph{Forschungszentrum {K}arlsruhe {R}eport {FZKA}} \textbf{6019}, 90
  (1998).

\bibitem[Huege et~al.(to be published)]{Huegearena2012}
T.~Huege, M.~Ludwig, and C.~James, \enquote{Simulating radio emission from air
  showers with {CoREAS},} in \emph{Proceedings of the ARENA 2012 workshop
  (Erlangen, Germany)}, AIP Conference Proceedings, to be published.

\end{thebibliography}
\end{document}